\documentclass[aps,prb,reprint,showpacs,flushbottom]{revtex4-1}
\usepackage{amsfonts}
\usepackage{amsmath}
\usepackage{amssymb}
\usepackage{graphicx}
\usepackage[dvips,colorlinks=true,bookmarks=false,citecolor=blue]{hyperref}
\usepackage{times}
\usepackage{CJK}

\input{tcilatex}
\begin{document}

\title{Anomalous spin Hall effects in Dresselhaus (110) quantum wells}
\author{Ming-Hao Liu}
\altaffiliation[Current address: ]{Institut f\"{u}r Theoretische Physik, Universit\"{a}t Regensburg, D-93040 Regensburg, Germany}%
\email{minghao.liu.taiwan@gmail.com}%
\affiliation{Department of Physics, National Taiwan University, Taipei 10617, Taiwan}%
\author{Ching-Ray Chang}
\email{crchang@phys.ntu.edu.tw}%
\affiliation{Department of Physics, National Taiwan University, Taipei 10617, Taiwan}%
\pacs{72.25.Dc, 71.70.Ej, 73.23.Ad}

\begin{abstract}
Anomalous spin Hall effects that belong to the intrinsic type in
Dresselhaus (110) quantum wells are discussed. For the out-of-plane
spin component, antisymmetric current-induced spin polarization
induces opposite spin Hall accumulation, even though there is no
spin-orbit force due to Dresselhaus (110) coupling. A surprising
feature of this spin Hall induction is that the spin accumulation
sign does not change upon bias reversal. Contribution to the spin
Hall accumulation from the spin Hall induction and the spin
deviation due to intrinsic spin-orbit force as well as extrinsic
spin scattering, can be straightforwardly distinguished simply by
reversing the bias. For the inplane component, inclusion of a weak
Rashba coupling leads to a new type of $S_y$ intrinsic spin Hall
effect solely due to spin-orbit-force-driven spin separation.
\end{abstract}

\date{\today}
\maketitle

\section{Introduction}

The intensive efforts on spin Hall effect (SHE) both experimentally and
theoretically during the past decade have successfully built another
milestone in condensed matter physics. Spin separation in semiconductors is
not only possible but natural, so that manipulating spin properties of
charge carriers in electronics is promising. The earliest theoretical idea
that up and down spins may laterally separate upon transport due to
asymmetric scattering was proposed in 1971.\cite{Dyakonov1971a,Dyakonov1971b}
More than three decades later, the power of optical measurements on high
quality mesoscopic samples made SHE in semiconductors no longer an idea but
an experimental fact.\cite{Kato2004} Right before the first observation of
Ref.~\onlinecite{Kato2004} in 2004, mechanisms of SHE was further extended
from spin-dependent scattering that was later categorized as extrinsic, to
spin-orbit-coupled band structure that was later categorized as intrinsic.%
\cite{Murakami2003,Sinova2004} Experimentally, most observations so far have
been attributed to the extrinsic SHE,\cite%
{Kato2004,Sih2005,Valenzuela2006,Seki2008} while evidence of the intrinsic
SHE\cite{Wunderlich2005} is relatively few. Nonetheless, intrinsic SHE
remains an important issue that until now still receives enduring efforts.%
\cite{Brune2010}

\begin{figure}[b]
\includegraphics[width=0.9\columnwidth,trim=0 0 0 0.5cm]{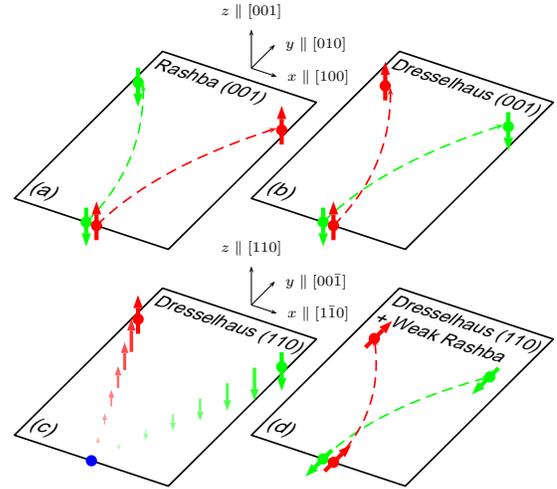}
\caption{(Color online) Spin deviation in $S_{z}$ due to spin-orbit force in
(a) Rashba and (b) linear Dresselhaus [001] systems. (c) Spin Hall induction
due to antisymmetric CISP along [00\={1}] in Dresselhaus [110] systems may
cause opposite $S_{z}$ accumulations as well, even though there is no
spin-orbit force. (d) In the case of linear Dresselhaus [110] plus weak
Rashba couplings, a coupled spin-orbit force given by Eq.\ \eqref{newFso}
may lead to spin deviation in $S_{y}$. }
\label{fig1}
\end{figure}

In the intrinsic SHE, spin separation is solely due to the underlying
spin-orbit coupling in the band structure, so that SHE can exist even in
systems free of scattering (but of finite sizes\cite{Inoue2004,Chalaev2005}%
). In the ballistic limit, the spin separation can be vividly visualized by
the transverse spin-orbit force \cite{Li2005,Nikolic2005} derived by using
the Heisenberg equation of motion,%
\begin{equation}
\mathbf{F}_{so}=\frac{m}{i\hbar }\left[ \frac{1}{i\hbar }\left[ \mathbf{r},%
\mathcal{H}\right] ,\mathcal{H}\right] ,  \label{Fso}
\end{equation}%
where $\mathbf{r}$ is the position operator and $\mathcal{H}$ is the
single-particle Hamiltonian. For well discussed two-dimensional systems with
Rashba coupling \cite{Bychkov1984} described by $\mathcal{H}_{R}=\left(
\alpha /\hbar \right) \left( p_{y}\sigma ^{x}-p_{x}\sigma ^{y}\right) $, as
well as linear Dresselhaus (001) coupling \cite{Dresselhaus1955,Dyakonov1986}%
\ described by $\mathcal{H}_{D}^{001}=\left( \beta /\hbar \right) \left(
p_{x}\sigma ^{x}-p_{y}\sigma ^{y}\right) ,$ the spin-orbit force is given by
\cite{Li2005}%
\begin{equation}
\mathbf{F}_{so}^{RD001}=\frac{2m\left( \alpha ^{2}-\beta ^{2}\right) }{\hbar
^{3}}\left( \mathbf{p}\times \mathbf{e}_{z}\right) \sigma ^{z},
\label{FsoRD001}
\end{equation}%
where $\mathbf{p}$ is the momentum and $\mathbf{e}_{z}$ is the unit vector
of the plane normal. Here $\alpha $ and $\beta $ are Rashba and Dresselhaus
coupling constants, respectively, and $\sigma ^{x},\sigma ^{y},\sigma ^{z}$
are Pauli matrices. Equation \eqref{FsoRD001} clearly depicts a lateral spin
deviation of the $S_{z}=(\hbar /2)\sigma ^{z}$ spin component with opposite
contributions from Rashba and Dresselhaus (001) couplings, as sketched in
Fig.\ \ref{fig1}(a) and (b), respectively.

SHE in Dresselhaus (110) quantum wells (QWs), on the other hand, is
relatively less discussed theoretically,\cite{Hankiewicz2006} although a
previous experimental effort \cite{Sih2005} had revealed in GaAs (110) QWs
the existence of SHE that was attributed to the extrinsic type. In this
paper, anomalous SHEs that belong to the intrinsic type in Dresselhaus (110)
QWs are discussed. We show that the spin Hall pattern of $S_{z}$ can be
induced when the transport direction is properly oriented, even though the
Dresselhaus (110) coupling does not result in spin-orbit force to separate
opposite $S_{z}$ spins upon transport [see Fig.\ \ref{fig1}(c)]. Moreover,
we propose a Rashba-coupling-assisted intrinsic SHE in $S_{y}$ that is truly
due to spin-orbit force under the interaction of Dresselhaus (110) plus a
weak Rashba couplings [Fig.\ \ref{fig1}(d)].

This paper is organized as follows. In Sec.~\ref{sec formula} we briefly
introduce the formulas required in the Landauer-Keldysh formalism employed
in the numerical analysis of Sec.~\ref{sec numerical}, where we visualize
the proposed spin Hall induction and Rashba-coupling-assisted SHE in $S_{y}$%
. Comparison of the present ballistic calculation with the diffusive
experiment of Ref.\ \onlinecite{Sih2005} will be discussed and the transport
parameters used in our numerical data will be remarked. We conclude in Sec.~%
\ref{sec conclusion}.

\section{Formulas\label{sec formula}}

\subsection{Linear Dresselhaus (110) coupling}

\begin{figure}[t]
\includegraphics[width=\columnwidth,trim=0 0 0.5cm 0]{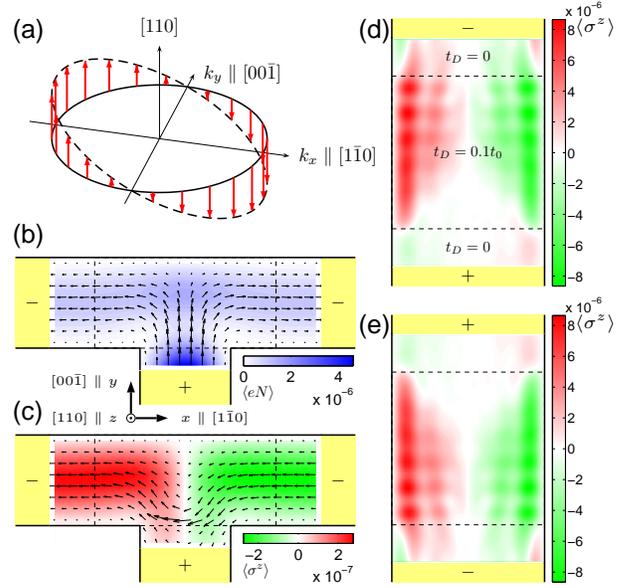}
\caption{(Color online) (a) Linear Dresselhaus [110] spin-orbit field in $k$
space. (b) Local charge density $\langle eN\rangle $ (color shading) and
local charge current density (arrows) in a $16a\times 8a$ Dresselhaus [110]
sample subject to three terminals with a weak bias $eV_{0}=10^{-3}t_{0}$.
(c) Local spin density $\langle \protect\sigma ^{z}\rangle $ (color shading)
and its corresponding local spin current density $\langle \mathbf{J}%
^{S_{z}}\rangle $ (arrows) in the same device as (b) with same conditions.
Spin Hall induction in a $40a\times 40a$ Dresselhaus [110] sample with (d)
upward bias and (e) downward bias; strong bias $eV_{0}=0.4t_{0}$ is applied.
Note that in (b)--(e), the regions outside the dashed lines are simulating
the leads (zero spin-orbit coupling and constant on-site energy set equal to
the applied bias).}
\label{fig2}
\end{figure}

The Dresselhaus (110) coupling up to the term linear in momentum can be
written as%
\begin{equation}
\mathcal{H}_{D}^{110}=-\frac{\beta }{\hbar }p_{x}\sigma ^{z},  \label{HD110}
\end{equation}%
where $x$, $y$, and $z$ axes are chosen along [1\={1}0], [00\={1}], and
[110], respectively. Throughout the present discussion, we will focus on
this Dresselhaus (110) linear term, so that without ambiguity we use the
same notation $\beta $ to denote its coupling strength. The spin-orbit field
subject to Eq.\ \eqref{HD110} is depicted in Fig.\ \ref{fig2}(a). Clearly,
when propagating with $\mathbf{k}=\left( \pm \left\vert k_{x}\right\vert
,k_{y}\right) $ electrons encounter antisymmetric spin-orbit fields on the
left and right sides of the [00\={1}] axis (or the $y$ axis). Hence the
current-induced spin polarization (CISP) effect \cite%
{Edelstein1990,Kato2004b,Liu2008a} is expected to build opposite $S_{z}$
spin densities at the two sides, as conceptually depicted in Fig.\ \ref{fig1}%
(c).

To better illustrate this spin Hall induction, we will in Sec.~\ref{sec
numerical} first consider a T-bar ballistic nanostructure, attached to left,
right, and bottom leads (from the top view) that are made of normal metals.
The central region is described by the square-lattice tight-binding
Hamiltonian,%
\begin{equation}
H=\left( U+4t_{0}\right) \openone\sum_{n}c_{n}^{\dag }c_{n}+\sum_{\langle
nm\rangle }c_{m}^{\dag }\mathbf{t}_{m\leftarrow n}c_{n},  \label{H}
\end{equation}%
where the sum over $\langle nm\rangle $ of the second term is run for the
sites nearest to each other, satisfying $\left\vert \mathbf{r}_{m}-\mathbf{r}%
_{n}\right\vert =a,$ $a$ being the lattice grid spacing and $\mathbf{r}_{n}$
the position vector of site $n$, and the hopping matrix is given by%
\begin{equation}
\mathbf{t}_{m\leftarrow n}=-t_{0}\openone-it_{D}d_{x}\sigma ^{z}.  \label{t}
\end{equation}%
Here $U$ is the on-site energy set to be constant over the whole sample, $%
t_{0}=\hbar ^{2}/2ma^{2}$ is the kinetic hopping energy, $\openone$ is the
identity, $c_{m}$ ($c_{m}^{\dag }$) annihilates (creates) an electron at
site $m$, $t_{D}=\beta /2a$ is the Dresselhaus hopping parameter, and $%
d_{x}=\left( \mathbf{r}_{m}-\mathbf{r}_{n}\right) \cdot \mathbf{e}_{x}$ is
the hopping displacement along $x$ from site $m$ to site $n$.

\subsection{Landauer-Keldysh formalism}

To image the nonequilibrium charge, charge current, spin, and spin current
densities under the influence of the biased leads, a powerful and convenient
approach is the Landauer-Keldysh formalism,\cite{Nikolic2005a,Datta1995}
especially for the present ballistic case free of particle-particle
interaction. In this formalism, physical quantities in a nonequilibrium but
steady state are expressed in terms of the lesser Green function matrix $%
G^{<}$, provided that those physical observables of interest are well
defined.\cite{Nikolic2006} Each matrix element $G_{mn}^{<}$ in our spin-1/2
electron system is a $2\times 2$ submatrix, so that the size of full $G^{<}$
amounts to $2N\times 2N$, $N$ being the total number of lattice grid points.
Following Ref.\ \onlinecite{Nikolic2006} with moderate extension, we have
and will use the local charge and spin densities%
\begin{eqnarray}
\langle eN_{n}\rangle  &=&\frac{e}{2\pi i}\int dE\func{Tr}_{s}G_{nn}^{<}
\label{<eN>} \\
\frac{\hbar }{2}\langle \sigma _{n}^{i}\rangle  &=&\frac{\hbar /2}{2\pi i}%
\int dE\func{Tr}_{s}\left[ \sigma ^{i}G_{nn}^{<}\right]   \label{<si>}
\end{eqnarray}%
for site $n$, and the local charge and spin current densities%
\begin{eqnarray}
\langle J_{n\rightarrow m}\rangle  &=&-\frac{e}{h}\int dE\func{Tr}_{s}\left[
\mathbf{t}_{m\leftarrow n}G_{nm}^{<}-\mathbf{t}_{n\leftarrow m}G_{mn}^{<}%
\right]   \label{<J>} \\
\langle J_{n\rightarrow m}^{S_{i}}\rangle  &=&-\frac{1}{8\pi }\int dE\func{Tr%
}_{s}[\left\{ \mathbf{t}_{m\leftarrow n},\sigma ^{i}\right\} G_{nm}^{<}
\notag \\
&&-\left\{ \mathbf{t}_{m\leftarrow n}^{\dag },\sigma ^{i}\right\} G_{mn}^{<}]
\label{<JSi>}
\end{eqnarray}%
for the flow from site $n$ to site $m$. Here $\func{Tr}_{s}$ is the trace
done in the spin space, the explicit energy $E$ dependence of $%
G_{mn}^{<}\left( E\right) $ is suppressed, and $\left\{ A,B\right\} =AB+BA$
is the anticommutator. For the present illustration of the spin Hall
induction, we will first consider a pure linear Dresselhaus (110) system and
use the hopping matrix \eqref{t} with $t_{D}=0.1t_{0}$, which is within a
reasonable range. Parameters extracted from experiments will be discussed
later in Sec.~\ref{sec para}. Other transport parameters are as follows:
hopping parameter $t_{0}=1,$ on-site energy $U=0$ (so that band bottom $%
E_{b}=0$), Fermi energy is $0.2t_{0}$ above $E_{b}$ (so that the square
lattice simply serves as the grid of a free electron gas). We will always
label $+$ and $-$ to indicate an applied bias voltage of $+eV_{0}/2$ and $%
-eV_{0}/2$ on each lead, respectively, with $eV_{0}>0$. (Note that $%
e=-\left\vert e\right\vert $ is the negative electron charge, and hence
electrons always flow from $+$ to $-$ signs). In the rest of our analysis,
we will focus on the nonequilibrium contribution \cite{Nikolic2006} of those
quantities listed in Eqs.\ \eqref{<eN>}--\eqref{<JSi>}, and hence the
integration range will be taken as $E_{F}-eV_{0}/2\rightarrow E_{F}+eV_{0}/2$%
.

\section{Numerical Analysis\label{sec numerical}}

\subsection{Spin Hall induction in $S_{z}$: Pure Dresselhaus (110) coupling}

Employing the Landauer-Keldysh formalism briefly introduced above, we now
drive electrons in the T-bar nanostructure from bottom to left and right
leads with $eV_{0}=10^{-3}t_{0}$, as shown in Fig.\ \ref{fig2}(b), where the
background color shading is determined by the local charge density $\langle
eN\rangle $ given in Eq.\ \eqref{<eN>}, while each arrow indicates the local
charge current density given by Eq.\ \eqref{<J>}. In Fig.\ \ref{fig2}(c),
the color shading is determined by the local spin $\langle S_{z}\rangle $
density [Eq.\ \eqref{<si>}] and clearly shows an antisymmetric $S_{z}$
polarization for electrons moving into the left and right leads because of
the opposite Dresselhaus (110) fields they feel. The local spin current
density indicated by the arrows therein is given by Eq.\ \eqref{<JSi>},
which is derived from the symmetrized spin current operator $%
J^{S_{i}}=\left\{ J_{m\rightarrow n}/e,S_{i}\right\} /2$ in a way similar to
Ref.\ \onlinecite{Nikolic2006}. A pure spin current from right to left is
observed; at right side the spin current is flowing toward left because of
the negative $S_{z}$ times the right moving particle current while at left
side the left flowing spin current stems from the product of the positive $%
S_{z}$ and the left moving particle current.

For a two-terminal device made of Dresselhaus (110) QW oriented along [00\={1%
}] ($y$ axis), spin\ Hall accumulation of opposite $S_{z}$ is therefore
expected, as shown in Fig.\ \ref{fig2}(d)--(e), where we consider a strong
bias voltage of $eV_{0}=0.4t_{0}$ for a $40a\times 40a$ sample. A striking
difference between the spin Hall induction introduced here and the spin Hall
deviation due to spin-orbit force is the independence of the accumulation
sign on the bias direction. Whether driving the electrons from bottom to top
[Fig.\ \ref{fig2}(d)] or from top to bottom [Fig.\ \ref{fig2}(e)], one
always observe a negative $S_{z}$ accumulation at right while positive at
left.

Contrary to the present ballistic nanostructure here, the SHE previously
observed in GaAs (110) QWs \cite{Sih2005} was in diffusive regime and
attributed to the extrinsic type. The experiment used ac lock-in detection
referenced to the frequency of a square wave alternating voltage with zero
dc bias offset, and the resulting signal is sensitive only to the difference
in Kerr rotation between positive and negative bias. Our spin Hall induction
that does not depend on bias direction may therefore either hardly
contribute or be subtracted in the result of Ref.\ \onlinecite{Sih2005}.

\begin{figure}[tb]
\includegraphics[width=\columnwidth]{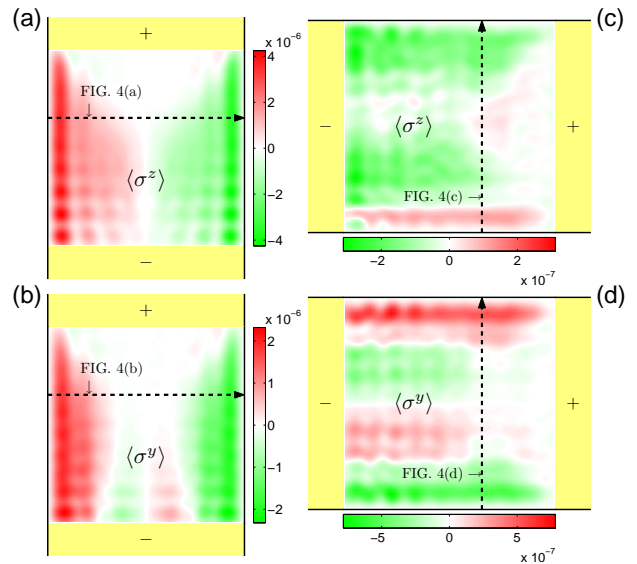}
\caption{(Color online) Imaging of local spin densities $\langle \protect%
\sigma ^{z}\rangle $ and $\langle \protect\sigma ^{y}\rangle $ in a $%
60a\times 60a$ [110] sample with (a)--(b) top-to-bottom and (c)--(d)
right-to-left orientations. A strong Dresselhaus [110] coupling $%
t_{D}=0.08t_{0}$ and a weak Rashba coupling $t_{R}=0.02t_{0}$ are used. Bias
is set $eV_{0}=0.4t_{0}$.}
\label{fig3}
\end{figure}

\subsection{Spin Hall deviation in $S_{y}$: Strong Dresselhaus (110) with
weak Rashba couplings\label{sec sy}}

Next we recall the spin-orbit force Eq.\ \eqref{Fso}. For pure Dresselhaus
(110) systems given by Eq.\ \eqref{HD110}, there is no way to obtain a
nonvanishing $\mathbf{F}_{so}$ since eventually the Pauli matrix $\sigma ^{z}
$ will commute with itself, even if the cubic term that is still in terms of
$\sigma ^{z}$ is involved. The only possibility in this case for a
nonvanishing $\mathbf{F}_{so}$ to survive is to introduce spin-orbit terms
involving $\sigma ^{x}$ or $\sigma ^{y}$. Combination of Rashba coupling
with the present linear Dresselhaus (110) term is therefore a natural
candidate, which is possible for, for example, asymmetric GaAs (110) QWs, as
are the cases of Ref.\ \onlinecite{Sih2005}. The spin-orbit force for this
Rashba-Dresselhaus (110) QW is%
\begin{equation}
\mathbf{F}_{so}^{RD110}=\frac{2m\alpha }{\hbar ^{3}}\left( \mathbf{p}\times
\mathbf{e}_{z}\right) \left( \alpha \sigma ^{z}-\beta \sigma ^{y}\right) .
\label{FsoRD110}
\end{equation}%
Without Rashba term $\alpha ,$ the spin-orbit force vanishes and zero spin
current is hence expected. From a gauge viewpoint, the existence of
equilibrium spin current in (110) QWs will require Rashba term to break the
pure gauge.\cite{Tokatly2010} Note also that the $\alpha $ squared
dependence for the $\sigma ^{z}$ component in Eq.\ \eqref{FsoRD110} is
similar to the result in Ref.\ \onlinecite{Sherman2005a}.

Here of particular interest is the case of weak Rashba coupling, such that
Eq.\ \eqref{FsoRD110} becomes

\begin{equation}
\left. \mathbf{F}_{so}^{RD110}\right\vert _{\alpha \ll \beta }\approx -\frac{%
2m\alpha \beta }{\hbar ^{3}}\left( \mathbf{p}\times \mathbf{e}_{z}\right)
\sigma ^{y},  \label{newFso}
\end{equation}%
which predicts a lateral spin Hall deviation in $S_{y}$ that requires a weak
but nonzero Rashba coupling $\alpha $. To further visualize the predicted
Rashba-assisted $S_{y}$ SHE, we consider a $60a\times 60a$ sample with
Dresselhaus (110) hopping $t_{D}=0.08t_{0}$ and Rashba hopping $t_{R}\equiv
\alpha /2a=0.02t_{0}$, attached to two leads under a bias voltage $%
eV_{0}=0.4t_{0}$. For the [001]-oriented (electron flow along $-y$) sample,
the $S_{z}$ spin Hall pattern due to spin Hall induction is observed in
Fig.\ \ref{fig3}(a). Meanwhile, an $S_{y}$ spin Hall pattern is also shown
in Fig.\ \ref{fig3}(b), which is a combined consequence of not only the spin
deviation Eq.\ \eqref{newFso} but also an antisymmetric CISP by the Rashba
coupling. Along the $-y$ axis, electrons with wave vector $\mathbf{k}=\left(
\pm \left\vert k_{x}\right\vert ,-\left\vert k_{y}\right\vert \right) $
encounter opposite $y$ component of the clockwise Rashba spin-orbit field:
negative for $+\left\vert k_{x}\right\vert $ and positive for $-\left\vert
k_{x}\right\vert $. Hence a spin Hall induction in $S_{y}$ due to Rashba
coupling contributes to Fig.\ \ref{fig3}(b) as well. In addition, the
contribution of the spin-orbit force Eq.\ \eqref{newFso} predicts a $+S_{y}$
($-S_{y}$) accumulation at left (right) side of the electron flow, for
lateral distance shorter than the spin precession length $L_{so}$ (around $%
15a$ here); the accumulation sign reverses when the lateral distance exceeds
$L_{so}$, as is the case in our $60a\times 60a$ here. Therefore the two
contributions, spin Hall induction and spin-orbit force Eq.\ \eqref{newFso},
are additive in Fig.\ \ref{fig3}(b).

For the [\={1}10]-oriented (electron flow along $-x$) sample, there is a
vague spin Hall pattern in $S_{z}$ because of the absence of the
antisymmetric CISP and weak spin-orbit force [Fig.\ \ref{fig3}(c)]. The
average of $\langle S_{z}\rangle $ over the whole sample basically reveals
the usual CISP effect as observed in Ref.\ \onlinecite{Sih2005}. The $S_{y}$
pattern, on the other hand, exhibits a clear spin Hall accumulation pattern
which is solely attributed to the spin-orbit force Eq.\ \eqref{newFso}, as
shown in Fig.\ \ref{fig3}(d). As explained, $-S_{y}$ ($+S_{y}$) accumulates
at left (right) side of the electron flow because the lateral distance has
exceeded $L_{so}$. Upon the bias reversal, the $\pm S_{y}$ edge
accumulations swap (not shown here), which is the general feature of the
spin Hall pattern due to spin deviation by intrinsic spin-orbit force, as
well as by extrinsic spin scattering. The spin Hall induction such as that
of $S_{z}$ in the Dresselhaus (110) case along $\pm y$, however, does not
have this feature. Another difference between the spin Hall patterns induced
by antisymmetric CISP and spin-orbit-driven spin deviation is that in the
former the signs of the spin accumulation do not change with the increasing
sample width, while in the latter they do. This difference can also be told
in Fig.\ \ref{fig3}: constant sign in each lateral side in panel (a) but
varying sign in panels (b) and (d).
\begin{figure}[tb]
\includegraphics[width=\columnwidth,trim=0 2.5cm 0 0]{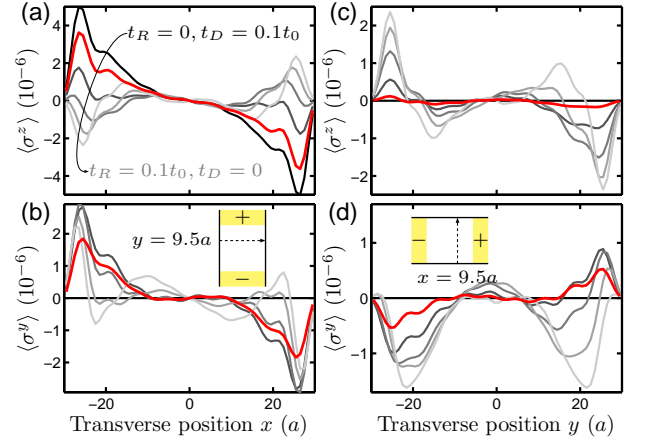}
\caption{(Color online) Local spin densities $\langle \protect\sigma %
^{z}\rangle $ and $\langle \protect\sigma ^{y}\rangle $ in a $60a\times 60a$
sample as a function of the transverse position with various coupling
parameters $t_{D}$ and $t_{R}$. The origin is set at the center of the
sample. As indicated in panel (a), the coupling parameters from black to the
lightest curves are $\left( t_{R},t_{D}\right) =\left( 0,0.1\right) t_{0},$ $%
\left( 0.02,0.08\right) t_{0},$ $\left( 0.04,0.06\right) t_{0},$ $\left(
0.06,0.04\right) t_{0},$ $\left( 0.08,0.02\right) t_{0},$ and $\left(
0.1,0\right) t_{0}$. In each panel, the red (dark gray) thick curve
correspond to Fig.\ \protect\ref{fig3}.}
\label{fig4}
\end{figure}

\subsection{From pure Dresselhaus (110) to pure Rashba cases}

Finally, we laterally scan the local spin densities $\langle S_{z}\rangle $
and $\langle S_{y}\rangle $ in Fig.\ \ref{fig4} at the positions marked by
the dashed lines in Fig.\ \ref{fig3}, for a set of various spin-orbit
coupling parameters from pure Dresselhaus (110) (black curves) to pure
Rashba (lightest gray curves). For the [001]-oriented sample, $S_{z}$ spin
Hall pattern shown in Fig.\ \ref{fig4}(a) gradually evolves from spin\ Hall
induction due to Dresselhaus (110) coupling to spin Hall deviation driven by
spin-orbit force due to Rashba coupling. In Fig.\ \ref{fig4}(b), the black
curve for the pure Dresselhaus (110) shows zero everywhere [and so are those
for Fig.\ \ref{fig4}(c)--(d)], while a weak Rashba coupling assists the
formation of the $S_{y}$ spin Hall pattern; the antisymmetric pattern holds
all the way to pure Rashba because the Rashba coupling contributes through
the antisymmetric CISP in this orientation as explained previously. For the [%
\={1}10]-oriented sample, turning on of the weak Rashba coupling builds $%
S_{y}$ spin Hall pattern [Fig.\ \ref{fig4}(d)] but not too much for $S_{z}$
[Fig.\ \ref{fig4}(c)]. Down to pure Rashba, the $S_{z}$ pattern recovers the
spin Hall accumulation due to spin-orbit force [Fig.\ \ref{fig4}(c)], while
that for $S_{y}$ shows symmetric CISP [Fig.\ \ref{fig4}(d)].

\subsection{Remark on transport parameters\label{sec para}}

In our numerical analysis for the pure Dresselhaus case, we have set $%
t_{D}/t_{0}=0.1$ mostly based on an illustrative reason. This coupling ratio
allows a direction comparison with Ref.\ \onlinecite{Nikolic2005a}, where $%
t_{R}/t_{0}=0.1$ is chosen, in the later part of coexisting Rashba and
linear Dresselhaus (110) couplings (such as Fig.\ \ref{fig4}).

Comparing with the GaAs (110) QWs used in the experiment of Ref.\ %
\onlinecite{Sih2005}, the coupling ratio $t_{D}/t_{0}$ may be one order
weaker than ours. The sample they used behaves like a single $75\unit{%
%TCIMACRO{\U{212b}}%
%BeginExpansion
\text{\AA}%
%EndExpansion
}$ Al$_{0.1}$Ga$_{0.9}$As QW. Using the relation $\beta =\gamma \langle
k_{z}^{2}\rangle $ with hard wall approximation $\langle k_{z}^{2}\rangle
\approx \left( \pi /w\right) ^{2}$ and $\gamma \approx 27\unit{eV}\unit{%
%TCIMACRO{\U{212b}}%
%BeginExpansion
\text{\AA}%
%EndExpansion
}^{3}$ for both GaAs and InAs QWs,\cite{Winkler2003} this well width of $w=75%
\unit{%
%TCIMACRO{\U{212b}}%
%BeginExpansion
\text{\AA}%
%EndExpansion
}$ leads to $\beta \approx 4.74\times 10^{-2}\unit{eV}\unit{%
%TCIMACRO{\U{212b}}%
%BeginExpansion
\text{\AA}%
%EndExpansion
}$. Effective mass was reported to be $m=0.074m_{0}$, $m_{0}$ the electron
rest mass. The sheet density is $n_{s}=1.9\times 10^{12}\unit{cm}^{-2},$
which allows us to estimate the location of the Fermi energy\cite{Datta1995}
$E_{F}-E_{b}=\pi \hbar ^{2}n_{s}/m\approx 6.\,\allowbreak 15\times 10^{-2}%
\unit{eV}$. In order for the long wavelength limit to be valid, the chosen
lattice constant $a$ has to yield a kinetic hopping constant $t_{0}$ that
keeps $E_{F}$ close to $E_{b}$. Choosing $a=2\unit{nm}$ leads to $%
t_{0}\approx 0.13\unit{eV}$ so that $E_{F}-E_{b}\approx 0.12t_{0}$ is
satisfying (recall $E_{F}-E_{b}=0.2t_{0}$ in our numerical results as well
as in Refs.\ \onlinecite{Nikolic2005a,Nikolic2006}). The coupling ratio with
this $a$ is $t_{D}/t_{0}\approx 0.01.$ The Rashba strength in Ref.\ %
\onlinecite{Sih2005} was reported to be $\alpha =0.018\unit{eV}\unit{%
%TCIMACRO{\U{212b}}%
%BeginExpansion
\text{\AA}%
%EndExpansion
}$, leading to $\alpha /\beta \approx 0.38$, which is not too far from our $%
t_{R}/t_{D}=0.25$ in Sec.~\ref{sec sy}. Replacing these parameters in our
results does not change significantly the main features we have shown.

Coupling ratio of $t_{D}/t_{0}=0.1$ is actually possible for QWs with
stronger Dresselhaus bulk coefficient $\gamma $. For InSb QWs,\cite%
{Winkler2003} we have $\gamma =760\unit{eV}\unit{%
%TCIMACRO{\U{212b}}%
%BeginExpansion
\text{\AA}%
%EndExpansion
}^{3}$. Consider the In$_{0.89}$Ga$_{0.11}$Sb QWs with effective mass $%
m=0.018m_{0}$ and sheet electron concentration $n_{s}=2.9\times 10^{11}\unit{%
cm}^{-2}$ reported in Ref.\ \onlinecite{Akabori2008}, where the QW width is
relatively thick: $30\unit{nm}$. If reducing the QW width to $7.5\unit{nm}$,
which is common in GaAs QWs, and assuming $a=3\unit{nm}$, the coupling ratio
is estimated as $t_{D}/t_{0}\approx 9.\,\allowbreak 45\times 10^{-2}$, close
to our $t_{D}/t_{0}=0.1$. The Fermi energy in units of $t_{0}$ is $%
(E_{F}-E_{b})/t_{0}=2\pi a^{2}n_{s}\approx 0.16$, which is also close to our
$(E_{F}-E_{b})/t_{0}=0.2$. Hence the transport parameters used in our
calculation are within a reasonable range.

\begin{table}[t]
\caption{Effective mass and Dresselhaus coefficients taken from Ref.\
\onlinecite{Winkler2003}.}
\label{tab1}%
\begin{ruledtabular}
\begin{tabular}{lllllll}
QW type & GaAs & AlAs & InAs & InSb & CdTe & ZnSe \\
\hline
$m/m_{0}$ & $0.067$ & $0.15$ & $0.023$ & $0.014$ & $0.09$ & $0.16$ \\
$\gamma (\unit{eV}\unit{%
%TCIMACRO{\U{212b}}%
%BeginExpansion
\text{\AA}%
%EndExpansion
}^{3})$ & $27.58$ & $18.53$ & $27.18$ & $760.1$ & $43.88$ & $14.29$%
\end{tabular}%
\end{ruledtabular}
\end{table}

\begin{figure}[t]
\includegraphics[width=0.9\columnwidth]{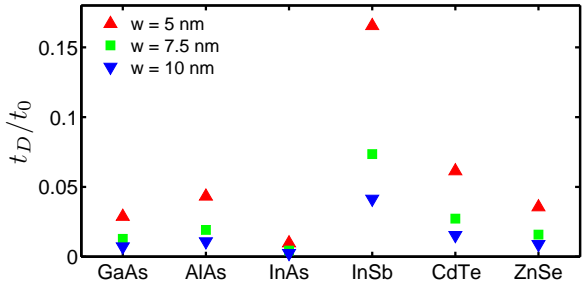}
\caption{(Online color) {}Coupling ratios $t_{D}/t_{0}$ estimated by Eq.\
\eqref{tD/t0} for various QWs with widths $w=5\unit{nm},7.5\unit{nm},10\unit{%
nm}$.}
\label{fig5}
\end{figure}

In general for a stronger $t_{D}/t_{0}$, which can be rewritten as%
\begin{equation}
\frac{t_{D}}{t_{0}}=\frac{\beta /2a}{\hbar ^{2}/2ma^{2}}\approx \frac{%
am\gamma }{\hbar ^{2}}\left( \frac{\pi }{w}\right) ^{2},  \label{tD/t0}
\end{equation}
a larger product $m\gamma $, and a thinner QW width $w$ will be required.
The effective mass $m$ and Dresselhaus coefficient $\gamma $ for various QWs
taken from Ref.\ \onlinecite{Winkler2003} are collected in Table \ref{tab1}.
For these QWs we use Eq.\ \eqref{tD/t0} with $a=3\unit{nm}$ to summarize the
coupling ratio $t_{D}/t_{0}$ in Fig.\ \ref{fig5} for QW widths $w=5\unit{nm}%
,7.5\unit{nm},10\unit{nm}$.

\section{Conclusion\label{sec conclusion}}

In conclusion, we have shown that spin Hall induction for $\pm \lbrack 001]$
transport in (110) QWs due to antisymmetric CISP of linear Dresselhaus
coupling that yields zero spin-orbit force is possible to generate a spin
Hall accumulation pattern in $S_{z}$, whose signs do not depend on the bias
direction. Experimental investigations with a dc bias offset in ballistic
(or at least quasi-ballistic) III-V (110) symmetric QWs may potentially
identify our proposed effect. From the coupling ratios summarized in Fig.\ %
\ref{fig5}, InSb (110) QW is promising for the presently proposed spin Hall
induction in $S_{z}$, while InAs is less suggested. A new type of spin Hall
deviation in $S_{y}$ is also predicted in the Dresselhaus (110) QWs in the
presence of a weak Rashba coupling. Experimental observation for this $S_{y}$
spin Hall effect may require a good control over the Rashba and Dresselhaus
couplings, which has been proved possible for (001) QWs,\cite%
{Ganichev2004,Meier2007,Koralek2009} and should be achievable also for (110)
QWs. We categorize these two intrinsic spin Hall mechanisms---spin Hall
induction in $S_{z}$ and spin Hall deviation in $S_{y}$, as anomalous SHEs.

We note that the Dresselhaus cubic term, neglected in the present study,
will become important when Fermi wave vector $k_{F}$ is long or QW width $w$
is thick. In this case the spin Hall induction in $S_{z}$ along $\pm $[001]
axis as discussed above should remain, while additional spin Hall induction
axes \emph{close to} [1\={1}1] and [\={1}12] will further emerge; see Fig.\
6.20 in Ref.\ \onlinecite{Winkler2003} and Fig.\ 2(a) in Ref.\ %
\onlinecite{Sih2005}. Inclusion of the Dresselhaus cubic term is left as a
future extending work.

\begin{acknowledgments}
We gratefully acknowledge V.\ Sih, R.\ Myers, Y.\ Kato, and D.\ Awschalom
for sharing their experimental insight. M.H.L.\ appreciates E.\ Ya.\ Sherman
for sharing his theoretical viewpoint. This work is supported by Republic of
China National Science Council Grant No. NSC 98-2112-M-002-012-MY3.
\end{acknowledgments}

\bibliographystyle{apsrev4-1}
\bibliography{mhl2}

\end{document}